 \documentclass[final,5p,times]{elsarticle}

\usepackage{amssymb}

\journal{Chaos, Solitons \& Fractals}

\begin{document}

\begin{frontmatter}



\title{Dynamical properties of a particle in a wave packet: scaling
invariance and boundary crisis}

\author{Diego F. M. Oliveira$^{\rm 1}$}
\ead{diegofregolente@gmail.com}
\author{Marko Robnik$^{\rm 1}$}
\ead{robnik@uni-mb.si}
\author{Edson D.\ Leonel$^{\rm 2}$}
\ead{edleonel@rc.unesp.br}
\address{$^{\rm 1}$CAMTP - Center For Applied Mathematics and
Theoretical Physics  University of Maribor - Krekova 2 - SI-2000 -
Maribor - Slovenia.\\
$^{\rm 2}$Departamento de Estat\'{\i}stica, Matem\'atica Aplicada e
Computa\c c\~ao - UNESP - Univ Estadual Paulista \\ Av. 24A,
1515 - Bela Vista - 13506-900 - Rio Claro - SP - Brazil.}

\begin{abstract}
Some dynamical properties present in a problem concerning the
acceleration of particles in a wave packet are studied. The dynamics of
the model is described in terms of a two-dimensional area preserving
map. We show that the phase space is mixed in the sense that there are
regular and chaotic regions coexisting. We use a connection with
the standard map in order to find the position of the first invariant
spanning curve which borders the chaotic sea. We find that the
position of the first invariant spanning curve increases as a power of
the control parameter with the exponent $2/3$. The
standard deviation of the kinetic energy of an ensemble of initial
conditions obeys a power law as a function of time, and saturates after
some crossover. Scaling formalism is used in order to characterize
the chaotic region close to the transition from integrability to
nonintegrability and a relationship between the power law exponents is
 derived. The formalism can be applied in many different systems
with mixed phase space. Then, dissipation is introduced into the model
and therefore the property of area preservation is broken, and consequently attractors are
observed. We show that after a small change of the dissipation,
the chaotic attractor as well as its basin of attraction are destroyed,
thus leading the system to experience a boundary crisis. The transient
after the crisis follows a power law with exponent $-2$.
\end{abstract}


\end{frontmatter}



\section{Introduction}

During the last decades many theoretical studies of area-preserving maps
have intensively been done \cite{ref1,ref2}. One of the most studied
models is the standard map proposed by B. V. Chirikov \cite{ref3a,ref3}
in 1969. The model  describes the motion of the kicked rotator. The
standard map can also be applied in different fields of science
including accelerator physics \cite{ref4}, plasma physics \cite{ref5}
and solid state physics \cite{ref6}. It has also been studied in
relation to problems of quantum mechanics and quantum chaos
\cite{ref7a,ref7b,ref7}, statistical mechanics \cite{ref8} and many
others.

The system we are considering in this paper is the dynamics of a
particle under the action of electrostatic waves. Usually, the model is
described using the formalism of discrete nonlinear maps. It is well
known that the structure of the phase space in such systems depends on
the combinations of both initial conditions and control parameters.
Basically they are classified in three different classes: (i)
integrable, (ii) ergodic and (iii) mixed. In case (i), the phase space
consists of invariant tori filling the entire phase space. In case (ii),
the ergodic systems have only one chaotic component, and the time
evolution of a single initial condition is enough to fill the phase
space \cite{ref10}-\cite{ref10c}. Finally, the case (iii) is most common
and an important property in the mixed phase space is that chaotic seas
are generally surrounding Kolmogorov-Arnold-Moser (KAM) islands which
are confined by a set of invariant curves. These mixed type systems are
subject of intense research in classical and quantum chaos
\cite{ref111}-\cite{ref1116}. The structure of the phase space of the
model we are considering is mixed and we derive analytically the
location of the invariant curves. We consider a connection with the
standard map close to the
transition from local to global chaos and obtain an effective control
parameter as well as the location of the first invariant torus bordering
the chaotic sea in the phase space. In this paper we study the scaling
behaviour of the model close to the transition from integrability to
nonintegrability \cite{ref11b} in a similar way as in \cite{ref11b4a}, 
as well as the dissipative version, and,
moreover, we analyze the boundary crisis. Our main concern is the
acceleration of the particle \cite{ref11b1,ref11b2,ref11b3} and
the saturation of the velocity \cite{ref11b4}. An extensive work using
scaling arguments was published recently in \cite{ref11b4a}.  However,
when the dissipative
dynamics is taken into account the structure of the phase space is
changed and attractors, namely, attracting fixed points and chaotic
attractor, might be observed. Then, increasing the strength of the
dissipation the edge of the basins of attraction of the attracting fixed
points and the edge of the basin of attraction of the chaotic attractor
(which corresponds to the stable and unstable manifolds of a saddle
fixed point) touch each other and as a result the chaotic attractor as
well as its basin of attraction is immediately destroyed. Such an event
is called boundary crisis \cite{refcr1,refcr2}. After the crisis, the
particle experiences a chaotic transient in the corresponding region
where a chaotic attractor existed prior to the crisis. The transient is
described by a power law with respect to the relative distance in the
control parameter where the crisis happens.

The paper is organized as follows. In section \ref{sec1} we construct
the two-dimensional map that describes the dynamics of the system and
discuss our numerical results. Conclusions  are drawn in section
\ref{sec2}.

\section{The model and numerical results}
\label{sec1}

We shall study the model of a point mass ($m$) charged ($e$)
particle moving in an electric field wave packet, as introduced by G.
Zaslavsky et al. \cite{ref11d}, which in general can be written as
\begin{equation} \label{eqR1}
\ddot{x} = \frac{e}{m} \sum_{n=-\infty}^{\infty} E_n \cos (k_nx -
\varpi_nt),
\end{equation}
where $E_n$ is the amplitude of the $n$-th Fourier component
of the electric field wave. They consider a wave packet with
the broad spectrum, and furthermore assume
$E_n=E_0$ for all $n$. Moreover, $\varpi_n \approx \varpi_0+ n \Delta \varpi$
and $k_n\approx k_0 + n \Delta k$. The group velocity of the wave packet
is thus $v_g=d\varpi/dk \approx \Delta \varpi/\Delta k$. If the
particle's velocity $v=\dot{x}$ is much larger than $v_g$, one
can use the approximation $\varpi_n \approx \varpi_0$.
Using all these assumptions we get
\begin{equation} \label{eqR2}
\ddot{x} = \frac{e}{m} E_0 \cos (k_0x-\varpi_0t)
\sum_{n=-\infty}^{\infty} \cos (n \Delta k x),
\end{equation}
and therefore using the Fourier decomposition of the periodic Dirac
delta function we have
\begin{equation} \label{eqR3}
\ddot{x} = \frac{e}{m} L E_0 \cos (k_0x-\varpi_0t)
\sum_{n=-\infty}^{\infty} \delta (x-nL),
\end{equation}
where we use $\Delta k= 2\pi/L$. Between the delta kicks we have free
motion of the particle and therefore the system of the two first order
ordinary differential equations for $\dot{x}$ and $\dot{v}$  emerging
from (\ref{eqR3}) can be formulated exactly as a discrete map. Defining
$\theta = k_0x-\varpi_0t$ and $\eta = mv|v|/2$ (kinetic energy),
and denoting by index $n$ their values just before the $n$-th kick,
we can write this map in the following form
\begin{equation}
P:\left\{\begin{array}{ll}
\eta_{n+1}=\eta_{n}+eE_0L\cos(\theta_n)~\\
\theta_{n+1}=\theta_n+k_0L~{\rm sign}{(\eta_{n+1})} -\varpi_0L \sqrt{m/2
\arrowvert \eta_{n+1} \arrowvert}
\end{array} 
\right.~,
\label{eq02}
\end{equation}
where $ {\rm sign(\eta_{n+1})} =1$ if $\eta_{n+1}>0$ or $\rm
sign(\eta_{n+1}) =-1$ if $\eta_{n+1}<0$. However, defining
$\phi=2\eta/m\varpi_0^2L^2$ and $\beta=\theta /2\pi-1/4$ as auxiliary
variables, and introducing a dissipation parameter $\sigma$, the map $P$
can be rewritten as
\begin{equation}
P:\left\{\begin{array}{ll}
\phi_{n+1}=(1-\sigma)\phi_{n}+\gamma\sin(2\pi\beta_n)~~\\
\beta_{n+1}=[\beta_n-{1 \over {2\pi|\phi_{n+1}|^{1/2}}} +\delta {\rm
sign}(\phi_{n+1})]~~~{\rm mod (1)}
\end{array} 
\right.~,
\label{eq03}
\end{equation}
where $\gamma=2eE_0/m\varpi_0^2L$, $\sigma \in [0,1]$ and
$\delta=k_0L/2\pi$. It is important to stress that $\delta$ is just a
shift of the phase and from now on we fix it as $\delta=0$ without loss
of generality. Here, $\gamma$ is the control parameter which controls
the transition from integrability ($\gamma=0$) to nonintegrability
($\gamma \ne 0$). As shown by Zaslavsky et al. \cite{ref11d} for large
$\gamma>>1$ we have adiabatic picture with intermittent but largely
regular behaviour. The determinant of the Jacobian matrix is ${\rm
det~J}=1-\sigma$. So for the case of $\sigma=0$ the mapping is area
preserving. Here we shall investigate some dynamical properties
close to the transition from integrability to nonintegrability assuming $\gamma<<1$.

\begin{figure}[t]
\centerline{\includegraphics[width=1.0\linewidth]{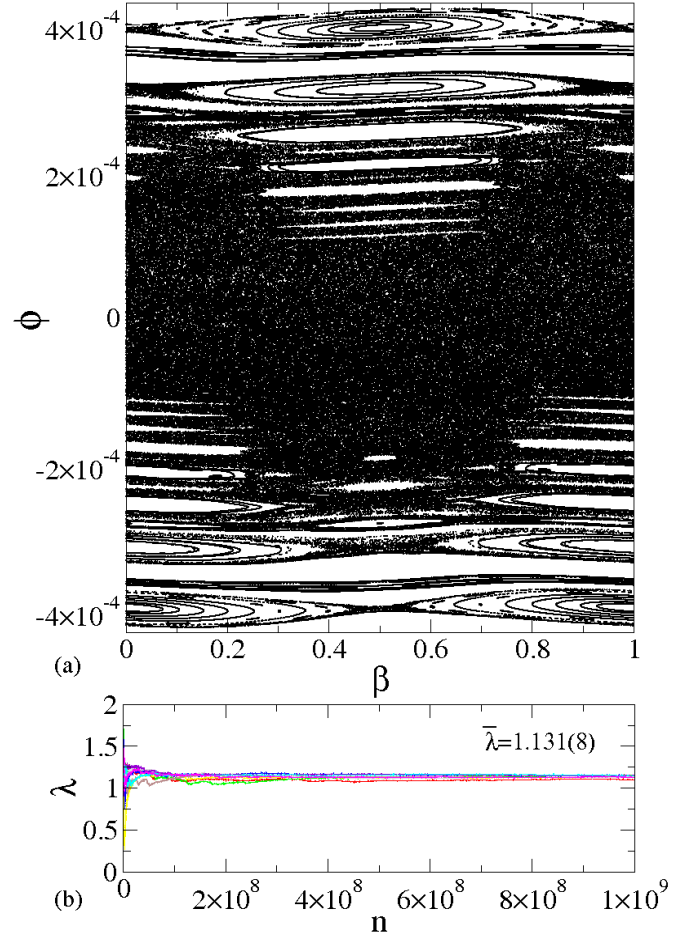}}
\caption{{(Color online) (a) Phase space generated from Eq. \ref{eq03}
the control parameter were $\gamma=10^{-5}$; (b) Convergence of the
positive Lyapunov exponent.}}
\label{fig01}
\end{figure}

Let us first consider the conservative case where $\sigma=0$. Figure
\ref{fig01} (a) shows the phase space generated from Eq. (\ref{eq03})
where the control parameter used is $\gamma= 10^{-5}$. As one can see
the phase space is mixed in the sense there are regular regions,
invariant spanning curves (invariant tori) and KAM islands coexisting
with a chaotic sea.  The
Lyapunov exponent is an important tool that can be used to classify
orbits as chaotic or not. As discussed in \cite{ref12}, the Lyapunov
exponents are defined as
\begin{equation}
\lambda_j=\lim_{n\rightarrow\infty}{1\over{n}}\ln|\Lambda_j|~~,~~j=1,
2~~,
\label{eq4}
\end{equation}
where $\Lambda_j$ are the eigenvalues of
$M=\prod_{i=1}^nJ_i(\beta_i,\phi_i)$ and $J_i$ is the Jacobian matrix
evaluated over the orbit $(\beta_i,\phi_i)$. However, a direct
implementation of a computational algorithm to evaluate Eq. (\ref{eq4})
has a severe limitation to obtain $M$. Even in the limit of short $n$,
the components of $M$ can assume very different orders of magnitude for
chaotic orbits and periodic attractors, yielding impracticable the
implementation of the algorithm. In order to avoid such problem, we note
that $J$ can be written as $J=\Theta T$ where $\Theta$ is an orthogonal
matrix and $T$ is a right up triangular matrix. Thus we rewrite $M$ as
$M=J_nJ_{n-1}\ldots J_2\Theta_1\Theta_1^{-1}J_1$, where
$T_1=\Theta_1^{-1}J_1$. A product of $J_2\Theta_1$ defines a new
$J_2^{\prime}$. In a next step, it is easy to show that
$M=J_nJ_{n-1}\ldots J_3\Theta_2\Theta_2^{-1}J_2^{\prime}T_1$. The same
procedure can be used to obtain $T_2=\Theta_2^{-1}J_2^{\prime}$ and so
on. Using this procedure, the problem is reduced to evaluate the
diagonal elements of $T_i:T_{11}^i,T_{22}^i$. Finally, the Lyapunov
exponents are given by
\begin{equation}
\lambda_j=\lim_{n\rightarrow\infty}{1\over{n}}\sum_{i=1}^n
\ln|T_{jj}^i|~~,~~j=1,2~~.
\label{eq005}
\end{equation}
If at least one of the $\lambda_j$ is positive then the system is
classified as chaotic. Figure \ref{fig01} (b) shows the behaviour of the
positive Lyapunov exponent. We have used $10$ different initial
conditions iterated  in the large chaotic region up to $10^{9}$ times. The average of the positive
Lyapunov exponent for our ensemble is $\bar{\lambda}=1.131(8)$, where
$0.008$ corresponds to the standard deviation of the 10 samples.

The structure of the phase space of the discrete dynamical system we
are dealing with and defined by the map $P$ [Eq. (\ref{eq03})] is
illustrated in Fig. \ref{fig01} (a). We can make a connection between
the standard map and the map $P$ in order to  find the position of the
first invariant spanning curve bordering the chaotic region and also to
characterize the transition from integrability to nonintegrability as
it has been done in \cite{ref2,ref12a}. The standard map was first
proposed by J. B. Taylor \cite{ref13} in order to study the existence of
the invariants of motion in magnetic traps. Later, B. V. Chirikov
\cite{ref3a,ref3} proposed another way to obtain the map and since then
it became clear that situations described by the  Chirikov map occur in
many physical systems. The Chirikov map is
\begin{equation}
C:\left\{\begin{array}{ll}
I_{n+1}=I_n+K\sin\left(\Phi_{n}\right)~\\
\Phi_{n+1}=\Phi_{n}+I_{n+1}~
\end{array}
\right.~,
\label{eq04}
\end{equation}
where $K$ is the control parameter (kick parameter). $K$ controls the transition from integrability, ($K=0$), to
nonintegrability, ($K \ne 0$), and also controls the transition from
local chaos $K<K_c$, where there still exists a set of invariant
spanning curves separating different regions in the phase space, to
global chaos, $K>K_c$, where all the global invariant curves are
destroyed. The critical $K$ is $K_c=0.971635\ldots$. The connection of
this result with the present model (\ref{eq03}) is as follows. Suppose
that close to the first invariant spanning curve, $\phi_{n}$ can be
written as
\begin{equation}
\phi_{n}\cong\phi^*+\Delta\phi_{n}~
\label{eq1}
\end{equation}
where $\phi^*$ is a typical value along the invariant spanning curve and
$\Delta\phi_{n}$ is a small perturbation of $\phi$. After defining $X_n =
2\pi\beta_n$, the second equation in the map (\ref{eq03}) can be written
as
\begin{equation}
X_{n+1}=X_n-{1 \over {|\phi_{n+1}|^{1/2}}}~.
\label{eq2}
\end{equation}
Using Eq. (\ref{eq1}), we can rewrite Eq. (\ref{eq2}) as
\begin{equation}
X_{n+1}=X_n-{1 \over {|\phi^*|^{1/2}}}\left[ 1+{{\Delta\phi_{n+1}} \over
{|\phi^*|^{1/2}}} \right]^{-1/2}~.
\label{eq3}
\end{equation}
Expanding Eq. (\ref{eq3}) in Taylor series and taking into account only
terms of first order, we can rewrite Eq. (\ref{eq3}) as
\begin{equation}
X_{n+1}=X_{n}+ {1 \over {|\phi^*|^{1/2}}} \left[ 1-{\Delta\phi_{n+1}
\over 2|\phi^*|}\right]~.
\label{eq5}
\end{equation}
Using Eq. (\ref{eq1}), the first equation of the map (\ref{eq03}) can
also be written as
\begin{equation}
\phi^*+\Delta\phi_{n+1}=\phi^*+\Delta\phi_{n}+\gamma\sin(2\pi\beta_n)~.
\label{eq6}
\end{equation}
Multiplying both sides of Eq. (\ref{eq6}) by ${1 \over 2|\phi^*| }$ and
adding $-{1 \over |\phi^*|^{1/2}} $, we define
\begin{equation}
I_{n}=-{ 1 \over |\phi^*|^{1/2}} + {{\Delta\phi_{n} \over 2|\phi^*|^{3/2}}}~,
\label{eq7}
\end{equation}
and re-write the map (\ref{eq03}) as
\begin{figure}[t]
\centerline{\includegraphics[width=1.0\linewidth]{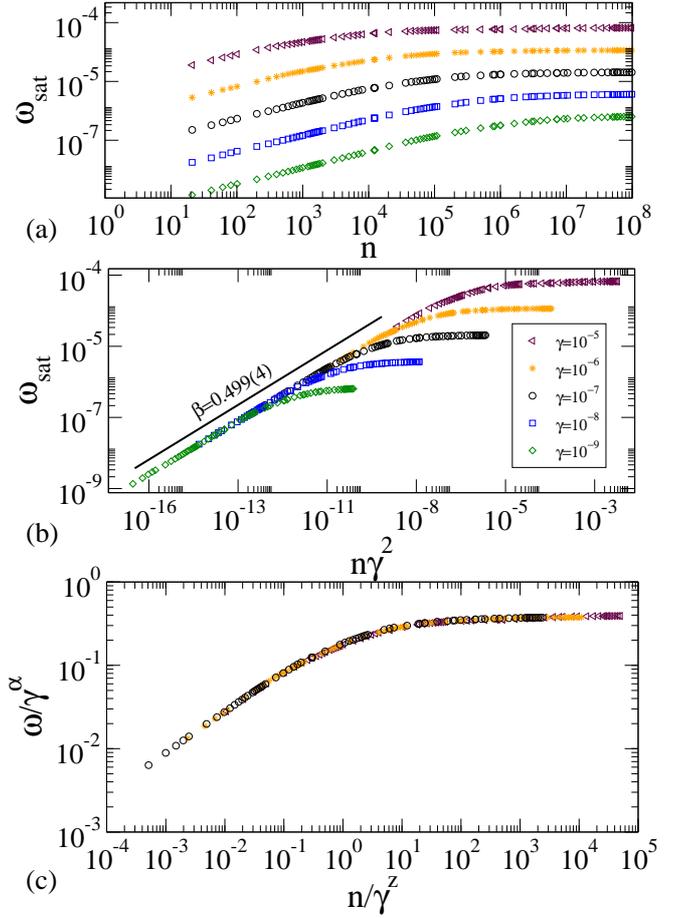}}
\caption{(Color online) (a) Behaviour of the standard deviation as
function of $n$ for different values of the control parameter $\gamma$.
(b) Their initial collapse after the transformation $n\gamma^2$. (c)
Their collapse onto a single and universal plot. The different curves are coded by the symbols in the legend in the same order (top-down).}
\label{fig33}
\end{figure}
\begin{figure}[htb]
\centerline{\includegraphics[width=1.0\linewidth]{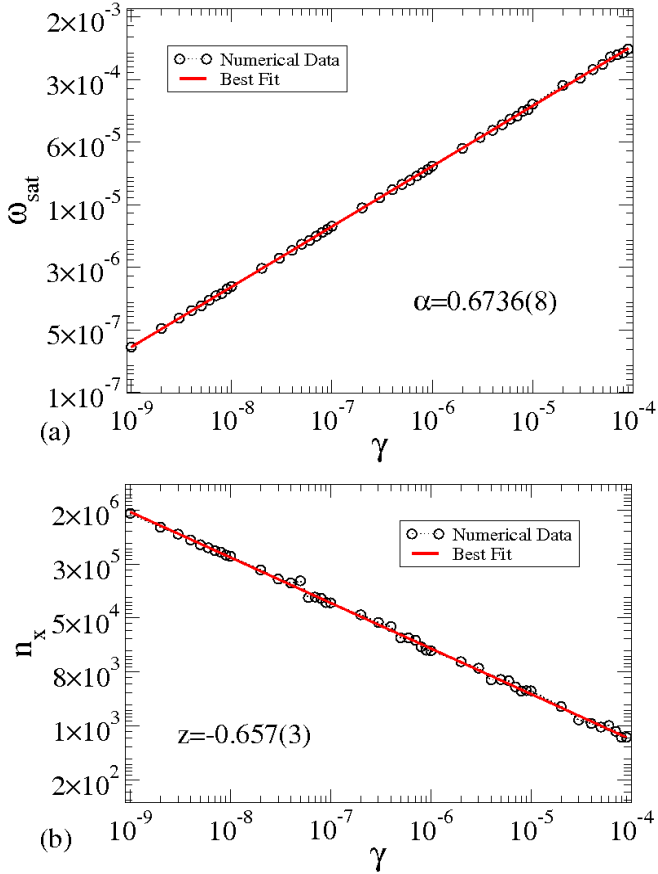}}
\caption{(Color online) (a) Plot of $\omega_{sat}$ as function of the
control parameter $\gamma$. (b) Behaviour of the crossover number $n_x$
against $\gamma$.}
\label{fig34}
\end{figure}

\begin{equation}
S:\left\{\begin{array}{ll}
I_{n+1}=I_n+{\gamma \over 2|\phi^*|^{3/2}}  \sin(X_{n}) ~ \\
X_{n+1}=X_{n}+I_{n+1}
\end{array}
\right.~,
\label{eq8}
\end{equation}
Comparing the standard map [see Eq. (\ref{eq04})] and the map $S$, we
see that there is an effective control parameter $K_{eff}$ which is
given by
\begin{equation} 
K_{eff}\cong{\gamma \over 2|\phi^*|^{3/2}}~.
\label{eq9}
\end{equation}
Since the transition from local to global chaos occurs at $K_{eff} >
0.971635\ldots$, the location of the first invariant spanning curve is
given by
\begin{equation}
|\phi^*|\cong {\left( \gamma \over 2 \times 0.971635\ldots\right)^{2/3}
} ~.
\label{eq10}
\end{equation}
According to Eq. (\ref{eq10}), the position of the first invariant
spanning curve changes with exponent $2/3$ as the control parameter
$\gamma$ varies. We conclude that for a given $\gamma$ a large chaotic sea is bordered by
 $\mid\phi\mid \leq \mid\phi^*\mid$. For example, for $\gamma=10^{-5}$, as in
Fig. \ref{fig01}(a), we see indeed that $\phi^*\cong 2.9 \times10^{-4}$
in perfect agreement with Eq. (\ref{eq10}). Similarly as in the standard
map in such regime $I^2$ grows diffusively, i.e. $\overline{I^2} \propto
n$ \cite{ref18}.

\begin{figure*}[t]
\begin{center}
\centerline{\includegraphics[width=9.0cm,height=8.cm]{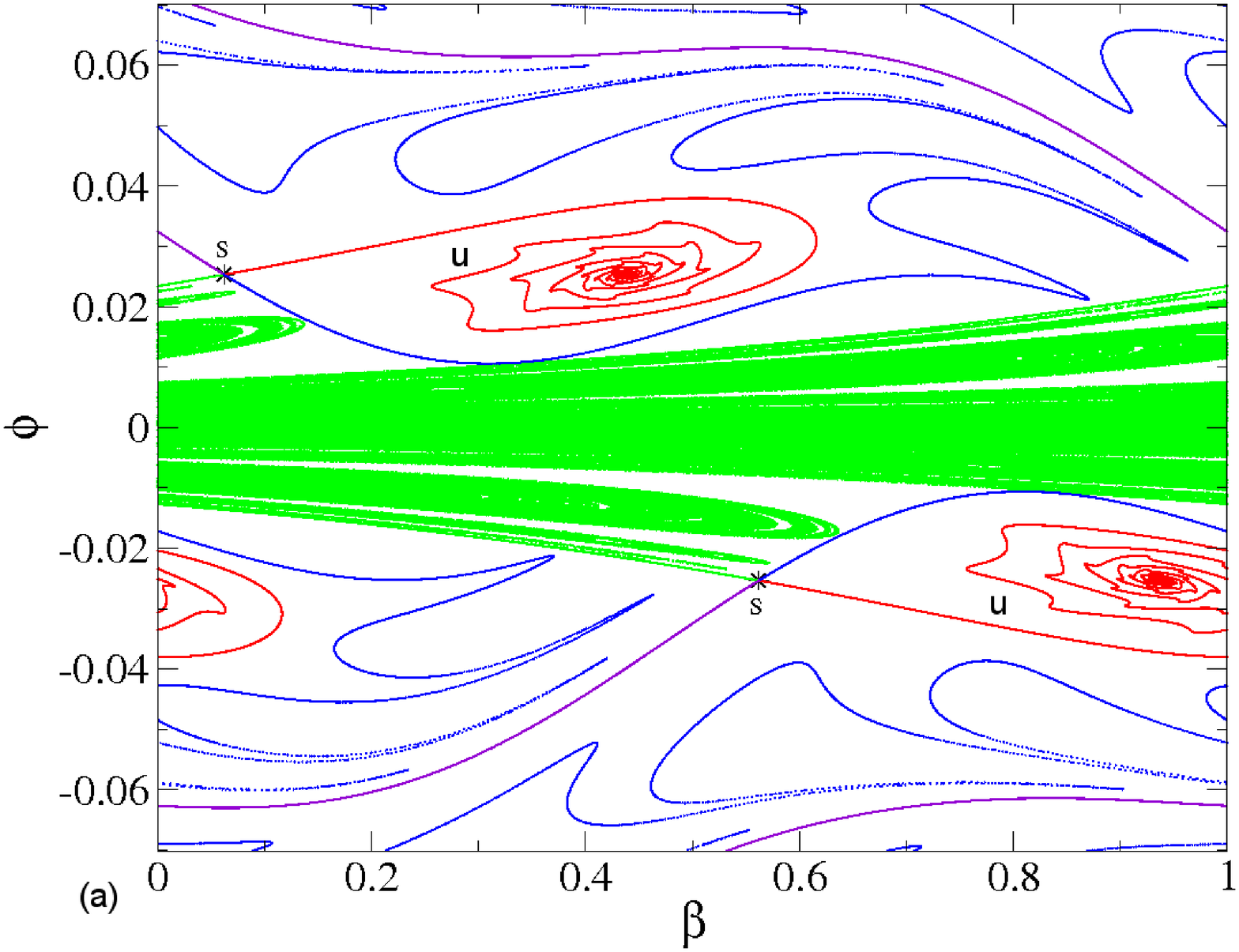}
            \includegraphics[width=9.0cm,height=8.cm]{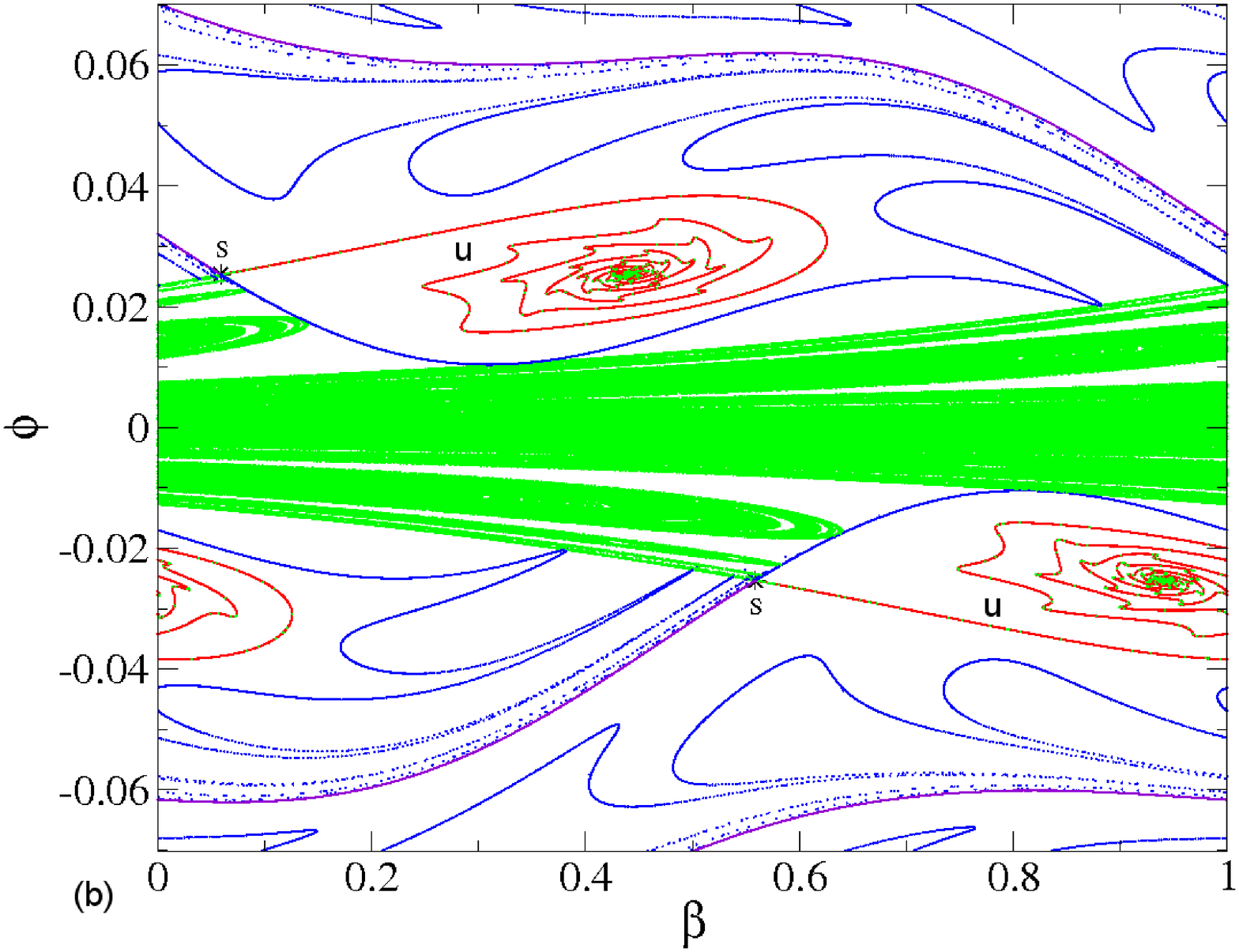}}
\end{center}
\caption{(Color online) Characterization of a boundary crisis. The control 
parameters used were $\gamma=10^{-2}$ and: (a) 
$\sigma=0.15$ (immediately before the crisis); (b) $\sigma=0.143$ 
(immediately after the crisis).}
\label{fig2}
\end{figure*}

We now assume small values for the parameter $\gamma$ and study the
average standard deviation of $\phi$ (kinetic energy), which is defined
as
\begin{equation}
\omega(n,\gamma)={{1}\over{C}}\sum_{i=1}^{C}\sqrt{\overline{\phi_i^2}
(n,\gamma)-{\overline{\phi_i}}^2(n,\gamma)}~,
\label{eq18}
\end{equation}
where
\begin{equation}
\overline{\phi}(n,\gamma)={{1}\over{n+1}}\sum_{i=0}^n \phi_i~.
\label{eq17}
\end{equation}
We have iterated Eq. (\ref{eq18}) up to $n=10^{8}$ for an ensemble of
$C=1000$ different initial conditions. The variable $\phi_0$ was fixed
as $\phi_0=10^{-2}\gamma$ and $\beta_0$ was uniformly distributed on
$[0,1]$. Figure \ref{fig33} shows the behaviour of $\omega(n,\gamma)$ as
function of $n$ for five different values of the control parameter
$\gamma$, as labelled in the figure. It is easy to see in Fig.
\ref{fig33}(a) two different kinds of behaviour. For short $n$,
$\omega(n,\gamma)$ grows  according to a power law and suddenly it bends
towards a regime of saturation for long enough values of $n$. The
crossover from growth to the saturation is marked by a crossover
iteration number $n_x$, which is very well defined by the intersection
of the acceleration and saturation straight lines. It must be emphasized
that different values of the parameter $\gamma$ generate different
behaviors for short $n$. However, applying the transformation
$n\rightarrow n\gamma^2$ all the curves start growing together for short
$n$, as shown in Fig. \ref{fig33}(b). Based on the behaviour shown in
Fig. \ref{fig33}, we can suppose the following scaling hypotheses:
\begin{enumerate}
\item{
When $n\ll{n_x}$, $\omega(n,\gamma)$ grows according to
\begin{equation}
\omega(n\gamma^2,\gamma)\propto ({n\gamma^2})^{\beta}~,
\label{eq19}
\end{equation}
where the exponent $\beta$ is the acceleration exponent;
}
\item{
For a long number of iteration, $n\gg{n_x}$, $\omega(n,\gamma)$
approaches a regime of saturation which is described by
\begin{equation}
\omega_{sat}(n \gamma^2,\gamma)\propto \gamma^{\alpha}~,
\label{eq20}
\end{equation}
where the exponent $\alpha$ is the saturation exponent;
}
\item{
The crossover iteration number that marks the change from growth to the
saturation is written as
\begin{equation}
n_x\propto \gamma^{z}~,
\label{eq21}
\end{equation}
where $z$ is the crossover exponent.
}
\end{enumerate}
After considering these three initial suppositions, we describe
$\omega(n,\gamma)$ in terms of a generalized homogeneous function of the
type
\begin{equation}
\omega(n\gamma^2,\gamma)=\tau\omega(\tau^a{n\gamma^2},\tau^b{\gamma})~,
\label{eq22}
\end{equation}
where $\tau$ is the scaling factor, $a$ and $b$ are scaling exponents
that in principle must be related to $\alpha$, $\beta$ and  $z$.
Since $\tau$ is a scaling factor we can choose $\tau^a{n\gamma^2}=1$, or
$\tau=(n\gamma^2)^{-1/a}$. Thus, Eq. (\ref{eq22}) is rewritten as
\begin{equation}
\omega(n\gamma^2,\gamma)=({n\gamma^2})^{-1/a}\omega_1[(n\gamma^2)^{-b/a}\gamma]~,
\label{eq23}
\end{equation}
where the function
$\omega_1[(n\gamma^2)^{-b/a}\gamma]=\omega[1,(n\gamma^2)^{-b/a}\gamma]$
is assumed to be constant for $n\ll{n_x}$. Comparing Eqs. (\ref{eq19})
and (\ref{eq23}), gives us $\beta=-1/a$. Choosing now $\tau^b\gamma=1$,
we have $\tau=\gamma^{-1/b}$ and Eq. (\ref{eq22}) is given by
\begin{equation}
\omega(n\gamma^2,\gamma)=\gamma^{-1/b}\omega_2(\gamma^{-a/b}n\gamma^2)~,
\label{eq24}
\end{equation}
where $\omega_2(\gamma^{-a/b}n\gamma^2)=\omega(\gamma^{-a/b}
n\gamma^2,1)$. It is also assumed as constant for $n\gg{n_x}$.
Comparison of Eqs. (\ref{eq20}) and (\ref{eq24}) gives us $\alpha=-1/b$.
Given the two different expressions of the saturation value $(n_x\gamma^2)^\beta \propto \gamma^\alpha$, we obtain 
\begin{equation}
n_x \propto \gamma^{{\alpha\over\beta}-2}~.
\label{eq25}
\end{equation}
Thus comparing Eq. (\ref{eq25}) and Eq. (\ref{eq21}), the crossover
exponent is $z={{\alpha/\beta}-2}$. Note that the scaling exponents are
determined if the critical exponents $\beta$ and $\alpha$ were
numerically obtained. The exponent $\beta$ is obtained from a power law
fitting for $\omega(n\gamma^2,\gamma)$ curves for the parameter
$\gamma\in[10^{-9},10^{-5}]$ for short iteration number, $n<<n_x$. Thus,
the average of these values gives us $\beta=0.499(4)\cong {1/2}$. 
Figure \ref{fig34} shows the behaviour of (a) $\omega_{sat} vs. \gamma$
and (b) $n_x vs. \gamma$. Applying power law fitting in the figure, we
obtain $\alpha=0.6736(8)\cong 2/3$ and $z=-0.657(3)$. We can also obtain
the exponent $z$ evaluating $z={{\alpha/\beta}-2}$ and the previous
values of both $\alpha$ and $\beta$. We found that $z=-0.651(9)$. This
result indeed agrees with our numerical data. This set of critical
exponents makes this transition to belongs to the same class of
universality as the phase transition observed in the problem of a
classical particle confined inside an infinitely deep box of potential
that has an oscillating square well \cite{square} or time varying barrier
in the middle \cite{barrier}.

Finally, in order to confirm our initial hypotheses and, since the
values of the critical exponents $\alpha$, $\beta$ and $z$ are now
already known, we can use the scaling laws. Fig. \ref{fig33}(c) shows
five different curves of $\omega$ generated from different values of the
control parameter $\gamma$ overlapped onto a single universal plot.

\begin{figure*}[t]
\begin{center}
\centerline{\includegraphics[width=9.0cm,height=8.cm]{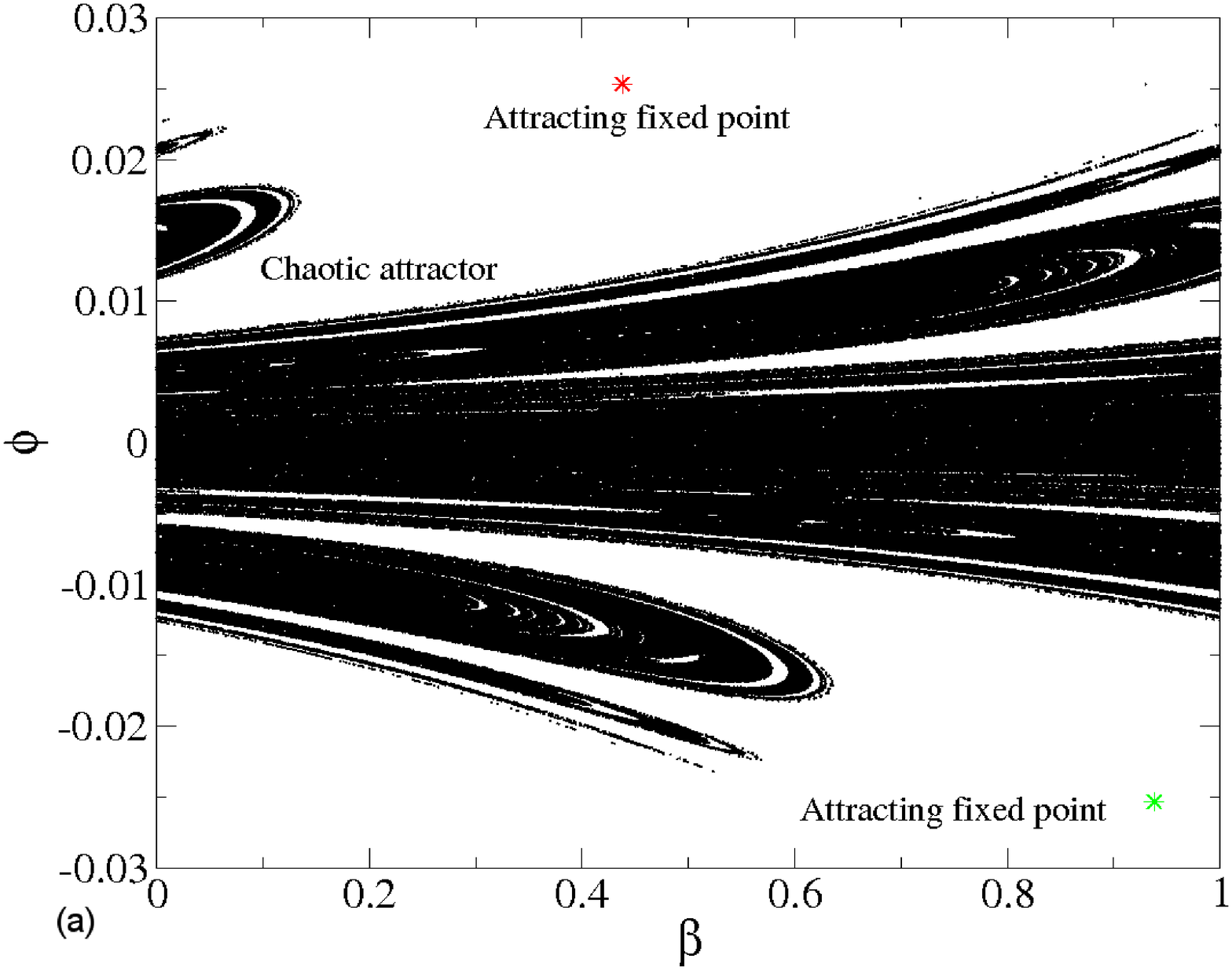}
            \includegraphics[width=9.0cm,height=8.cm]{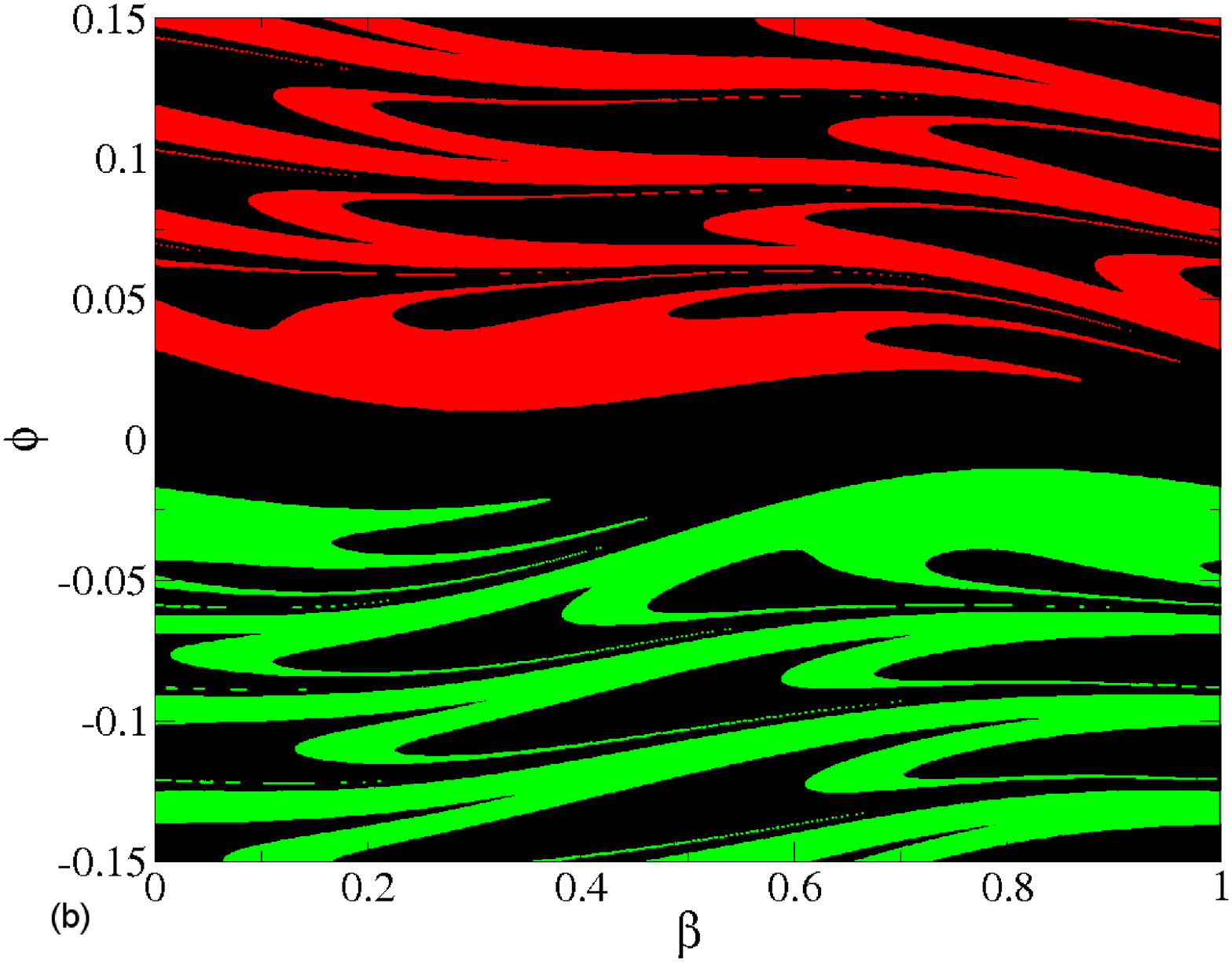}}
\end{center}
\caption{(Color online) (a) Attracting fixed points for $m=-1$
represented by a upperward red star, downward green star and a chaotic attractor shown in black. (b) Their
corresponding basin of attraction. The control parameters used in (a)
and (b) were $\gamma=10^{-2}$ and $\sigma=0.15$. The region in black is
the basin of attraction of the chaotic attractor and the region in red
color is the basin for the upper attracting fixed point and the region in green corresponds to the basin of
attraction of the lower attracting fixed point.}
\label{fig3}
\end{figure*}

Let us now present our results for the dissipative dynamics i.e.
$\sigma\ne 0$. It is well known that when dissipation is considered, the
mixed structure of the phase space is changed and elliptical fixed
points can in principle turn into attracting fixed points (sinks) and
chaotic seas might be replaced by chaotic attractors. Figure \ref{fig2}
shows the corresponding stable and unstable manifolds for two different
saddle fixed points which are obtained by solving $\phi_{n+1}=\phi_n$
and $\beta_{n+1}=\beta_n+m$. The two solutions of these equations of map \ref{eq03}  are
\begin{enumerate}
\item{\begin{eqnarray}
\phi=\left[{{-1} \over {2m\pi}}\right]^2~,~\beta^{(1)}={1 \over
2\pi}{\rm \arcsin}{\left({\phi\sigma \over \gamma}\right)}~~,
\label{eq10a}
\end{eqnarray}
}
and $\beta^{(2)}=1/2-\beta^{(1)}$, for each integer  $m=\pm 1,\pm 2,\pm3,\ldots$, and
\item{\begin{eqnarray}
\phi=-\left[{{-1} \over
{2m\pi}}\right]^2~,~\beta^{(1)}={{1}\over{2}}-{1 \over 2\pi}{\rm
\arcsin}{\left({\phi\sigma \over \gamma}\right)}~~,
\label{eq10b}
\end{eqnarray}
and $\beta^{(2)}=3/2-\beta^{(1)}=1+(1/2\pi)\arcsin(\phi\sigma/\gamma)$, for each integer  $m=\pm 1,\pm 2,\pm3,\ldots$. and {\rm arcsin} is evaluated on
the fundamental domain $[-\pi/2,\pi/2]$ where it is uniquely defined.
}
\end{enumerate}
For positive $\phi$ and $m=\pm1$ in Eq. \ref{eq10a} we find that $\beta^{(1)}$ corresponds to the hyperbolic (saddle) fixed point,
whilst $\beta^{(2)}$ corresponds to the attracting fixed point, as is illustrated in Fig. \ref{fig2} (a). For the negative $\phi$ the same conclusion applies, respectively.
For the whole sequence of larger $\mid m \mid \geq 2$ we have a sequence of fixed points but they can not be seen.
The saddle fixed points given by Eq. (\ref{eq10a}-\ref{eq10b}) are marked by a star
and are shown in the upper part of Fig. \ref{fig2}(a) while the Eq.
(\ref{eq10b}) gives the saddle fixed point shown in Fig. \ref{fig2}(a)
in the lower part, also marked by a star. The two branches of the unstable manifolds of each
saddle are are obtained via iteration of the map $P$ with appropriate
initial conditions. The upper [Eq. (\ref{eq10a})] and lower [Eq.
(\ref{eq10b})] unstable branches are marked by $u$ and shown in red in Fig. \ref{fig2}. They
converge to the attracting fixed points, while the downward [Eq.(\ref{eq10a})] and upperward [Eq. (\ref{eq10b})] branch generate the
chaotic attractor and are shown in green. On the other hand, the
stable manifolds consist basically in trajectories heading towards
the saddle points and the construction of the stable manifolds is
slightly different from those of the unstable manifolds where we have to
known the inverse of the mapping $P$, which we denote by $P^{-1}$. The
basic procedure is that
$P^{-1}(\phi_{n+1},\beta_{n+1})=(\phi_n,\beta_n)$, thus the map $P^{-1}$
is given by
\begin{equation}
P^{-1}:\left\{\begin{array}{ll}
\phi_{n}={\phi_{n+1}-\gamma\sin(2\pi\beta_n) \over (1-\sigma)}~\\
\beta_{n}=\beta_{n+1}+{1 \over {2\pi\mid \phi_{n+1} \mid^{1/2} }}
\end{array} 
\right.~.
\label{eq11}
\end{equation}

Once we have found $P^{-1}$, the stable manifolds, which corresponds
to the borders between the basins of attraction for the attracting fixed
points and the chaotic attractor are shown in blue and violet in Fig.
\ref{fig2}. By reducing the value of the dissipation parameter $\sigma$
 the unstable manifold of each saddle point touches simultaneously the stable
manifold and as a consequence the chaotic attractor as well as its basin
of attraction are immediately destroyed. Such an event is called as a
boundary crisis \cite{refcr3,refcr4,refcr5,refcr6} and happens
simultaneously for the manifolds arising from both Eqs. (\ref{eq10a})
and (\ref{eq10b}) as shown in Fig. \ref{fig2}(b) and is characterized
for the first time in this model. The control parameters used in Fig.
\ref{fig2} were $\gamma=10^{-2}$ and: (a) $\sigma=0.15$ (immediately
before the crisis); (b) $\sigma=0.143$ (immediately after the crisis).

\begin{figure}[t]
\centerline{\includegraphics[width=1.0\linewidth]{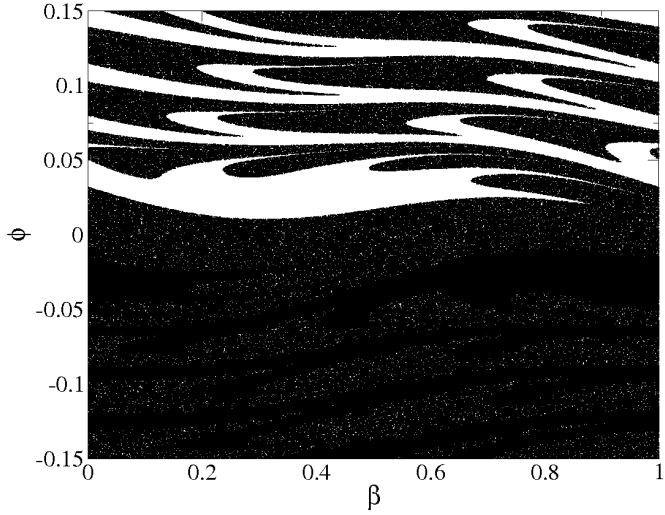}}
\caption{(Color online) Basins of attraction of the two sinks observed
in the phase after the boundary crisis. The interwoven structure is
evident. The control parameters used were $\gamma=10^{-2}$ and
$\sigma=0.143$.}
\label{Fig_bacias}
\end{figure}

Before the crisis and for the combination of control parameters used in
Fig. \ref{fig2}(a) ($\gamma=10^{-2}$ and $\sigma=0.15$), there are three
different attractors, namely: (i) two attracting fixed points related
to $m=\pm1$ (the central regions where the red spirals are converging to)
and (ii) a chaotic attractor [see Fig. \ref{fig3} (a)]. One might
expect that there must be three different basins of attraction. This is
indeed true as one can see in Fig. \ref{fig3}(b). The procedure used to
obtain the basins of attraction for the chaotic and attracting fixed
points consist basically in iterating a grid of initial conditions in the
plane $\beta\times\phi$ and look at their asymptotic behavior. We have
used a range for the initial $\beta$ as $\beta\in[0,1]$ and
$\phi\in[-0.15,0.15]$. Both ranges of $\beta$ and $\phi$ were divided in
$1000$ parts each, leading to a total of $10^6$ different initial
conditions and for the values of the control parameters that we have
considered, each combination of $(\beta,\phi)$ was iterated up to $10^5$
times. After the crisis, the time evolution of an initial condition in
the corresponding region of the chaotic attractor before the crisis
leads the orbit to make an incursion towards the region of the,
now, chaotic transient until being captured by one of
the two sinks. The structure of the two basins of attraction belonging to the two 
attracting fixed points is very complex and it is shown in Fig. \ref{Fig_bacias}, so that  it is difficult to decide which initial condition
will go to which of the two attractors, as there are some obviously some regions in the phase spcae where both basins of attraction are dense.

\begin{figure}[t]
\centerline{\includegraphics[width=1.0\linewidth]{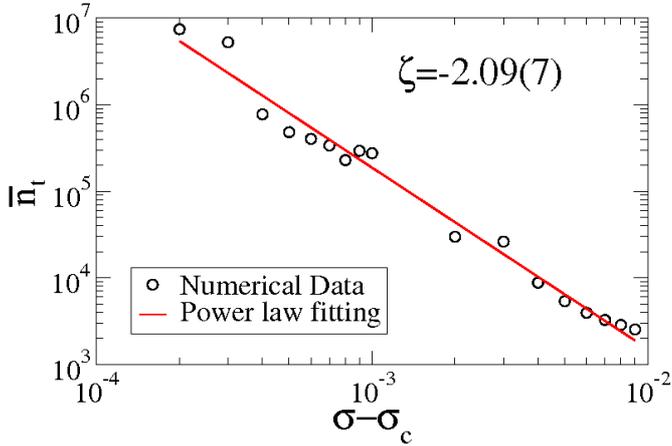}}
\caption{(Color online) Plot of $n_{t}~{\rm vs.} \mu$ for a fixed
$\gamma=0.01$. A power law fitting gives $\zeta=-2.09(7)$.}
\label{Fig_trans}
\end{figure}

We may therefore argue that an initial condition chosen in the region of
the phase space that produced a chaotic dynamics before the crisis, now gives rise to a chaotic transient. We denote as transient the
corresponding number of iterations that the orbit typically spends until it finds
the appropriate route to one of the attracting
fixed points. The transient is described by a
power law of the type
\begin{equation}
n_{t}\propto \mu^\zeta~,
\label{transiente}
\end{equation}
where $\mu=\sigma-\sigma_c$ with $\sigma_c<\sigma$. For a fixed
$\gamma=0.01$, our numerical simulations lead to the critical value of
$\sigma_c=0.143056\ldots$. Figure \ref{Fig_trans}
shows the behaviour of the average transient plotted as a function of
$\mu$ considering a fixed $\gamma=0.01$. A power law fitting gives the
exponent $\zeta=-2.09(7)$. The procedure used to simulate the transient
consists in evolving in time an ensemble of $B=5\times 10^3$ different
initial conditions randomly chosen along the region of the chaotic attractor as observed just before the
crisis, and calculated the time $n_{t}^i$, where $i$ is a member of the
ensemble, that the particle spends until reaches one of the two attracting
fixed points. The average transient is calculated as
\begin{equation}
\bar{n_{t}}={{1}\over{B}}\sum_{i=1}^Bn_{t}^i~,
\label{trans}
\end{equation}
where $i$ specifies an orbit of the ensemble. The exponent
$\zeta=-2.09(7)$ has, within an uncertainty error, the same value as the
exponent obtained for the boundary crisis observed in a dissipative
Fermi-Ulam model \cite{refcr4}. The Fermi-Ulam model consists of a
classical particle that is confined to bounce between two rigid walls.
One of them is fixed and the other one is periodically time-varying.
Despite the remarkable difference of the two models (the main difference being the power of the term in the denominator of the two mappings, which is
 $1$ in the Fermi-Ulam model and  $1/2$ in our model), the chaotic transient after the boundary crisis
marking the converging to the attracting fixed points follows the same
rule in both models and is given by the same exponent $\cong -2$.

\section{Conclusions}
\label{sec2}

We have studied the problem of a charged particle in the electric field
of a wave packet. We have confirmed through analytical arguments that
close to the invariant spanning curves the dynamics of our model can be
described by the standard map. Such  approach allows us to find the
position of the first invariant spanning curve (chaos border) as a
function of the control parameter $\gamma$. Once we have found that the
position of the first invariant torus changes as a power of the control
parameter $\gamma$ with the exponent $2/3$, we have studied the
behaviour of the chaotic sea close to the transition from integrability
to nonintegrability (small values of the control parameter $\gamma$)
using scaling arguments. We have shown that there exists an analytical
relationship between the critical exponents, namely, acceleration
exponent, saturation exponent and crossover exponent. Our scaling
hypotheses have been confirmed by the perfect collapse of all curves
onto a single universal plot. Finally, for the dissipative dynamics we
have found the expressions for the saddle fixed points and constructed
the corresponding unstable and stable manifolds. We have shown that
increasing the dissipation, the unstable manifold touches the stable
manifold and the chaotic attractor as well as its basin of attraction
are completely destroyed. We have shown that such a destruction is
caused by a boundary crisis. Finally the transient marking the approach
to the sink is characterized by a power law with exponent $\cong -2$.

\section*{Acknowledgment}

D.F.M.O. acknowledges the financial support by the Slovenian Human Resources Development and Scholarship Fund (Ad futura Foundation). M. R. acknowledges the financial support by The Slovenian Research Agency (ARRS).
E. D. L. is grateful to FAPESP, CNPq and FUNDUNESP, Brazilian agencies.


\end{document}